\documentclass[fleqn]{aa}  

\usepackage{graphicx}
\usepackage{xcolor}
\usepackage{hyperref}
\usepackage{txfonts}
\usepackage{calligra}
\usepackage{calrsfs}
\usepackage[T1]{fontenc}
\usepackage[mathcal]{eucal}
\usepackage{longtable}
\newcommand{\Msun}{\, {\rm M}_{\odot}}

\begin{document}

%   \title{Mapping dark matter on milliarcsecond scales with lensed stars}

    \title{A high-resolution view of the source-plane magnification near cluster caustics in wave dark matter models.}
  
  \titlerunning{HR wave DM}
  \authorrunning{Diego et al.}

   \author{J. M. Diego
       \inst{1}\fnmsep\thanks{jdiego@ifca.unican.es}
       \and Alfred Amruth  \inst{2}
       \and Jose M. Palencia \inst{1}
       \and Tom Broadhurst \inst{3,4,5}
       \and Sung Kei Li \inst{2} % Keith   
       \and Jeremy Lim \inst{2}  
       \and Rogier A. Windhorst \inst{6}
       \and Adi Zitrin \inst{7}
       \and Alexei V. Filippenko \inst{8}   
       \and Liliya L. R. Williams \inst{9,10}      
        \and Ashish K. Meena \inst{7}
        \and Wenlei Chen \inst{11}
        \and Patrick L. Kelly \inst{9,10}
%       \and
%       \and
%       Tom J. Broadhurst \inst{4,5,6}
%       \and     
%       Patrick L. Kelly \inst{7,8}
%       \and
%       Liliya L. R. Williams \inst{7,8}
%       \and
%       Adi Zitrin \inst{3}
%       \and
%       Jeremy Lim \inst{2}   
%       \and
    }      
  \institute{Instituto de F\'isica de Cantabria (CSIC-UC). Avda. Los Castros s/n. 39005 Santander, Spain %1 Chema
%              \email{jdiego@ifca.unican.es}
        \and
        Department of Physics, The University of Hong Kong, Pokfulam Road, Hong Kong  %2 Amruth
        \and
          Department of Physics, University of the Basque Country UPV/EHU, E-48080 Bilbao, Spain % 3 Tom
         \and
          DIPC, Basque Country UPV/EHU, E-48080 San Sebastian, Spain % 4 Tom
          \and
          Ikerbasque, Basque Foundation for Science, E-48011 Bilbao, Spain % 5 Tom
          \and 
          School of Earth and Space Exploration, Arizona State University, Tempe, AZ 85287-6004, USA % 6
          \and
         Physics Department, Ben-Gurion University of the Negev, P.O. Box 653, Be’er-Sheva 84105, Israel %7 Adi
         \and 
        Department of Astronomy, University of California, Berkeley, CA 94720-3411, USA % 8 Alexei  
        \and
         Minnesota Institute for Astrophysics, University of Minnesota, 116 Church Street SE, Minneapolis, MN 55455, USA % 9 Liliya, Pat
         \and 
         School of Physics and Astronomy, University of Minnesota, 116 Church Street, Minneapolis, MN 55455, USA  % 10 Liliya, Pat
         \and
         Department of Physics, Oklahoma State University, 145 Physical Sciences Bldg, Stillwater, OK 74078, USA % 11 Wenlei
         }
%         \and        
%        \and
%         Physics Department, Ben-Gurion University of the Negev, P.O. Box 653, Be’er-Sheva 84105, Israel %3 Ashish
%         \and        
%          Department of Physics, University of the Basque Country UPV/EHU, E-48080 Bilbao, Spain % 4 Tom
%         \and
%          DIPC, Basque Country UPV/EHU, E-48080 San Sebastian, Spain % 5 Tom
%          \and
%          Ikerbasque, Basque Foundation for Science, E-48011 Bilbao, Spain % 6 Tom
%         \and
%         Minnesota Institute for Astrophysics, University of Minnesota, 116 Church Street SE, Minneapolis, MN 55455, USA % 7 Pat, Liliya
%         \and 
%         School of Physics and Astronomy, University of Minnesota, 116 Church Street, Minneapolis, MN 55455, USA  % 8 Pat, Liliya
%         \and 
%          }
%   \date{Received September 15, 1996; accepted March 16, 1997}

 \abstract{
We present the highest resolution images to date of caustics formed by wave dark matter ($\psi$DM) fluctuations near the critical curves of cluster gravitational lenses. We describe the basic magnification features of $\psi$DM in the source plane at high macromodel magnification and discuss specific differences between the $\psi$DM and standard cold dark matter (CDM) models. The unique generation of demagnified counterimages formed outside the Einstein radius for $\psi$DM is highlighted. 
%We demonstrate that 
Substructure in CDM cannot generate such demagnified images of positive parity,  thus providing a definitive way to distinguish $\psi$DM from CDM. Highly magnified background sources with sizes  $r\approx 1pc$, or approximately a factor of ten  smaller than the expected de Broglie wavelength of $\psi$DM, offer the best possibility of discriminating  between $\psi$DM and CDM. These include objects  such as very compact stellar clusters at high redshift that JWST is finding in abundance.
   }
   \keywords{gravitational lensing -- dark matter -- cosmology
               }

   \maketitle
%
%-------------------------------------------------------------------

\section{Introduction}
%%%%%%%%%%%%%%%%%%%%%%%%
% Intro CDM
The nature of dark matter (DM, hereafter) is arguably one of the biggest mysteries in modern physics. The cold DM (CDM) model successfully explains the observations on large scales from galaxy surveys \citep{eBOSS2021,DES2022,DESI2024} to the cosmic microwave background \citep{PlanckCosmo2020}. From these observations, the standard cosmological model can be derived with percent precision in its most relevant parameters \citep{PlanckCosmo2016,PlanckCosmo2020}. On smaller scales, differences between predictions based on $N$-body simulations of CDM and observations have been used as examples that challenge the CDM model, but  these discrepancies might be due to insufficient modelling of the complex baryon physics in $N$-body simulations \citep{Weinberg2015}. Perhaps the biggest challenge to CDM resides in the lack of direct confirmation from any of the multiple experiments on Earth. In particular, the continuing search for weakly interacting massive particles (WIMPs) is quickly exhausting the still available space of parameters where long-sought WIMPs were originally expected to be found. This absence of direct confirmation has promoted the appearance of challengers to the CDM model. Among these, wave DM \citep[$\psi$DM, hereafter;][]{Press1990,Widrow1993,Peebles2000,Hu2000,Schive2016} has recently attracted attention as an alternative model that can simultaneously explain the large-scale structure of the Universe, the alleged small-scale departures from the CDM expectations, and the lack of success in direct search experiments.

%CDM alternatives
There are other scenarios in which dark matter physics, or various baryonic effects \citep[e.g.,][]{Ragagnin2024}, alter the properties of halos. For example, warm DM models predict less substructure on subgalactic scales. Surviving halos in warm DM have lower concentrations than their CDM counterparts, and therefore have a suppressed lensing efficiency, decreasing the contribution from millilenses to the lensing probability. On the other hand, self-interacting DM can cause halos to undergo core collapse, a process that dramatically raises their central density, potentially to a degree that causes them to become supercritical for lensing \citep{Gilman21}. Alternatively, $\psi$DM is expected to increase the magnification in regions relatively far from cluster critical curves (CCs), by producing smaller CCs around the largest $\psi$DM fluctuations.

%Intro waveDM 
In $\psi$DM, the DM particle is incredibly light ($\sim 15$ orders of magnitude smaller than the mass of the classic axion in quantum chromodynamics, for a axion-photon coupling constant of $\sim 10^{-17}$ GeV$^{-1}$). In this work, we assume the DM reaches the Bose Einstein Condensate. In this case, the DM mass is the only free parameter, and the self-interaction and ALP decay can be neglected.
The small mass in $\psi$DM results in distinctive behavior of the DM fluctuations on small scales ($<1$ kpc), while on Mpc scales, both $\psi$DM and CDM predict similar structures. With such a small mass, $\psi$DM behaves as a quantum fluid with a de Broglie wavelength on parsec (pc) to kiloparsec (kpc) scales. Interference of the $\psi$DM fluid results in constructive and destructive interference, where locally the density contrast can have variations of order unity, $\Delta\rho/\rho \approx 1$. When projecting these large density perturbations along the line of sight, the resulting fluctuations in the projected surface mass density are smaller but still present (by the central limit theorem). However, these small fluctuations are enhanced near the CCs of massive lenses and leave their mark in background objects through distortions in the magnification. These distortions extend over scales that are comparable to $\lambda_{\rm dB}$, the de Broglie wavelength of $\psi$DM. Hence, objects that have sizes comparable to $\lambda_{\rm dB}$ can experience differential magnification, depending on whether their counterimages are aligned with a region in the lens having slightly more or less interference.  This pervasive interference substructure causes the CCs to become corrugated on the de Broglie scale \citep{Chan2020,Laroche+22,Amruth2023,Liu2024}, and increasingly so for more massive halos, with many detached islands where the magnification diverges at relatively large offsets from the cluster CCs \citep{Amruth2023,Laroche+22,Powell2023}. 

$\psi$DM has been recently tested with observations of lensed objects, such as quasars \citep{Amruth2023,Powell2023}. The results are inconclusive, with some observations favoring $\psi$DM \citep{Amruth2023} while others disfavoring it \citep{Powell2023}.  New observations of distant objects with the James Webb Space Telescope (JWST) are providing a wealth of small but luminous objects that can in principle be used in similar studies to test DM models, including $\psi$DM. To perform these tests, a key issue is to understand the magnification properties in different models. For CDM (and its substructure) there is abundant literature exploring the magnification near the CCs of clusters. For $\psi$DM the literature is more scarce. 

One of the limitations of earlier work on magnification from $\psi$DM is the relatively low resolution that does not allow one to properly resolve the $\psi$-caustics in the source plane. Resolving the caustics from $\psi$DM is particularly challenging near the CCs of clusters, where the macromodel magnification is $\mu>100$. To simulate a region in the source plane of $100\times100$ pc$^2$, the corresponding area in the image plane needs to be over 100 times larger, with the pixel size sufficiently small in order to resolve the small caustics. 
%This is mostly due to the difficulties of simulation large scales (much larger than $\lambda_{dB}$), with sufficient resolution to resolve very small objects ($r<<1$ pc), and near regions of high macromodel magnification ($\mu_{macro}>100$) that require the area in the image plane to be at least  $\mu_{macro}>100$ times larger tan the target area in the source plane. 
The main focus of this paper is to address this issue and provide the first very-high-resolution view of the $\psi$DM caustics in the source plane, and to study the probability of magnification in the source plane of the $\psi$DM model.  

The paper is organized as follows.
In Section~\ref{Sec_wDM} we briefly introduce key concepts in $\psi$DM models. 
The simulations of the magnification produced by $\psi$DM are discussed in Section~\ref{Sec_wDM_Sims}. 
Section~\ref{Sec_CC} shows the resulting magnification (and its probability distribution) in the image plane and as a function of macromodel magnification.   Section~\ref{Sec_Caustics} then shows the corresponding magnification pattern and probability in the source plane. 
We discuss our results in Section~\ref{Sec_Discussion} and conclude in Section~\ref{Sec_Conclusions}. 
A flat cosmological model is adopted with $\Omega_m=0.3$ and H$_0=70$ km s$^{-1}$ Mpc$^{-1}$. For our simulations, we assume a lens at redshift $z=0.4$ and a source at $z=1$. Our results have a weak dependence on this choice.

\section{Wave Dark Matter}\label{Sec_wDM}
%%%%%%%%%%%%%%%%%%%%%%%%%%%%%%%%%%%%%%%%%%%

In this model, DM has density fluctuations at scales given by the de Broglie wavelength and the halo mass \citep{Schive2016}, from the dependence on momentum:
\begin{equation}
\lambda_{\rm dB}=15\,\left( \frac{10^{-22}\, {\rm eV}}{m_{\psi}} \right) \left( \frac{10^{15}\, \Msun}{M_{\rm cluster}}\right)^{1/3}\, \, {\rm pc}\, ,
\end{equation}
where $m_{\psi}$ is the mass of the ultralight axion-like particle (ALP). 
For ALP masses $m_{\psi} \approx 10^{-22}$\,eV and a $10^{15}\, \Msun$ cluster, this scale corresponds to 3\,mas in the lens plane (assuming $z_{\rm lens}\approx 0.4$).

Similarly to microlenses (typically stars in the intracluster medium), $\psi$DM fluctuations are ubiquitous across the lens plane, and as in the case of microlenses  and millilenses (globular clusters and dwarf galaxies in the cluster), these fluctuations get amplified near the CC by the macromodel. The one main difference between $\psi$DM and other small perturbations in the deflection field in the CDM scenario (mostly microlenses and millilenses) is that $\psi$DM perturbations can be either positive or negative in the convergence, while in CDM the perturbations in the mass are always positive.\footnote{In CDM, positive perturbations in mass in regions of the lens plane with negative parity are equivalent to negative perturbations in mass in the symmetric regions with positive parity.} This difference has profound implications for the resulting probability of magnification, as we discuss in other sections of this paper.

In this work we are interested in studying the caustics produced by $\psi$DM fluctuations. Before exploring a full simulation, it is interesting to study a simple toy model of an individual $\psi$DM fluctuation. These fluctuations are the result of projecting along the line of sight of the lens the individual granules with positive or negative underdensities ($\delta\rho/\rho \approx 1$). If the amplitude of the fluctuations of the individual granules can be described as a random Gaussian field with dispersion $\sigma_g$, their projection is also a Gaussian random field but with a dispersion that scales as $\sqrt{N_g}\sigma_g$, where $N_g$ is the number of granules that get projected along the line of sight. The mean projected signal (or mean of the Gaussian distribution) scales as $\sigma_s/\sqrt{N_g}$ by virtue of the central limit theorem and converges to the total projected mass, which is similar in both CDM and $\psi$DM, except in the very central region of the cluster where the $\psi$DM soliton produces cores \citep[see][]{Schive2016}. We are interested in regions near the cluster CCs, and hence far away from the soliton core.  The lensing deflection field  from the projected $\psi$DM granules is also a Gaussian field with fluctuations that can be well described by two-dimensional (2D) Gaussian functions with full width at half-maximum intensity (FWHM) comparable in size to the de Broglie wavelength. This configuration is typical of convergence fields with continuous positive and negative fluctuations (with respect to the mean) and a similar pattern for the shear (except this is rotated by 90 degrees with respect to the convergence field).

%%%%%%%%%%%%%%%%%%%%%%%%%%%%%%%%%%%%%%%%%%%%%%%%%%%%%%%%%%%
%Figure made by Dell /home/jdiego/Lensing/Microlensing/WaveDM_Amruth/Plot_Caustics_1Granule.pro
\begin{figure} 
   \includegraphics[width=9.0cm]{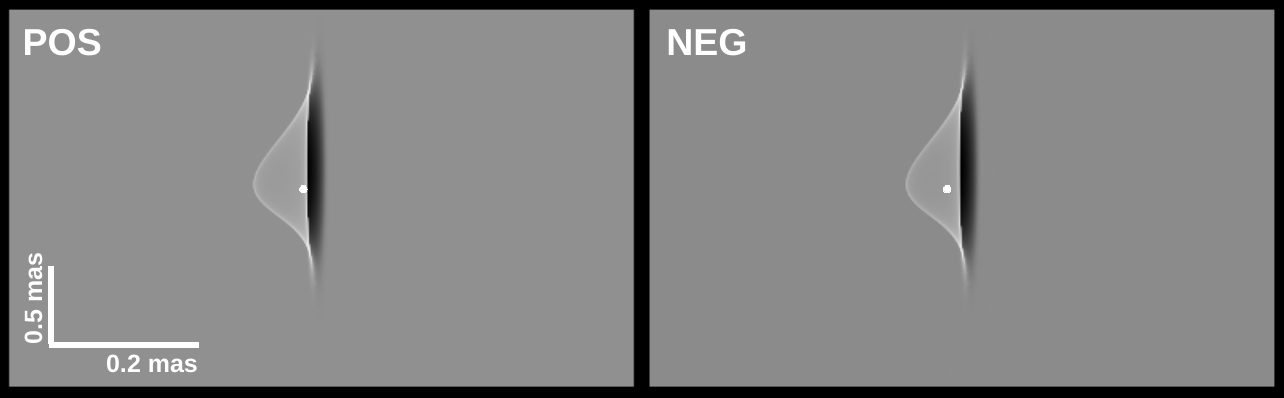}  
      \caption{Caustics for a toy model of an individual and isolated $\psi$DM perturbation. For this example, the $\psi$DM fluctuation is positive.
      The left and right panels show the caustics produced by the same 2\,pc scale $\psi$DM fluctuation but on the two sides of the lens CC, and at the same distance. That is, they represent the fluctuation in the positive- and negative-parity sides with the same macromodel magnification. The caustics are almost indistinguishable with just a small shift in position. The white dot marks the same position in the source plane, so the shift of the caustics with respect to this point indicates the shift in position when we vary the parity. A negative fluctuation in the deflection field of $\psi$DM (not shown) results in similar caustics, but mirrored in the horizontal direction.}
         \label{Fig_Caustics_ToyModel}
\end{figure}
%%%%%%%%%%%%%%%%%%%%%%%%%%%%%%%%%%%%%%%%%%%%%%%%%%%%%%%%%%%

We simulate one of these Gaussian projected fluctuations and embed it in a macromodel lensing potential at some small distance, $d$, from the cluster CC and with macromodel magnifications $\mu=\pm100\times1.6$. The two signs account for the two parities, and we fix the radial magnification to $\mu_r=1.6$ since this varies very slowly near cluster CCs. 
%On the other hand, the tangential component of the macromodel magnification, $\mu_t$, can vary rapidly near the CC, where at short distances, $d$, from the CC, it usually goes as $\mu_t(d)\propto d^{-1}$, diverging at the smooth macromodel CC.  
That is, we simulate the same $\psi$DM fluctuation on both sides of the cluster CC  but at a similar distance from the CC, so the macromodel magnification is the same in both parities (but with different sign).

For the $\psi$DM fluctuation we chose a small scale of just 2\,pc,  but this is irrelevant since for the discussion of this toy model  we are just interested in the morphology of the caustic, not on its scale. The resulting caustics are shown in Figure~\ref{Fig_Caustics_ToyModel} for the two parities. 

At first, the caustics appear identical for both parities, but on closer inspection we appreciate a very small shift in the position of the caustics. Ignoring this small shift (much smaller than the de Broglie wavelength), we can conclude that the magnification properties are identical on both sides of the lens CC. In earlier work, \cite{Amruth2023} found a preference for images with negative parity to be less magnified, but in that case the macromodel lens was a galaxy, not a cluster, where at short distances from the CC the magnification changes more rapidly with distance for images having negative parity. For clusters, the assumption that the magnification is the same at short distances ($d<1"$) from the CC is a much better approximation. 
This is a unique feature of $\psi$DM substructure, since any other substructure in CDM (positive with respect to the background by definition) results in significant differences in the probability of magnification between the sides with negative and positive parities \citep[see, e.g.,][]{Diego2018,Williams2024}. For instance, substructures in CDM in regions with negative parity can demagnify by a significant amount, rendering small objects that fall within the demagnification region undetectable. To compensate for the demagnification on the side with negative parity, the caustics around substructures in CDM are more powerful on the negative-parity side than on the positive-parity side.

On the contrary, CDM substructures cannot significantly demagnify background objects when these substructures are found in the region of the lens with positive parity, and their caustics are also less powerful than the caustics from the same lens but placed on the negative-parity side. 
These distinctive features are not present, or have different properties, in $\psi$DM, thus opening the door to test $\psi$DM models through the observation of multiply lensed objects in regions with positive and negative parity. 

For instance, we observe in Figure~\ref{Fig_Caustics_ToyModel}  that on the side with positive parity, demagnification is possible near the caustic, something that cannot be achieved by CDM substructures (see the discussion about this in Section~\ref{Sec_Discussion}). An observation of a small object ($\sim 1$\,pc) that produces two counterimages (one with positive parity and one with negative parity), such as a very small region with luminous stars like the compact stellar cluster R136, or the small but very luminous clusters found in lensed galaxies at high redshift \citep{Vanzella2023,AngelaNature2024,Messa2024}, can present anomalous demagnification in the counterimages found in regions with positive parity that would be very hard to explain with CDM.

%%%%%%%%%%%%%%%%%%%%%%%%%%%%%%%%%%%%%%%%%%%%%%%%%%%%%%%%%%%
%Figure made by Dell /home/jdiego/Lesnsing/Microlensing/WaveDM_Amruth/Plot_AlphaAmruth_AlphaGaussian.pro and AlphaXTrue_vs_AlphaXGauss.pro
\begin{figure} 
   \includegraphics[width=9.0cm]{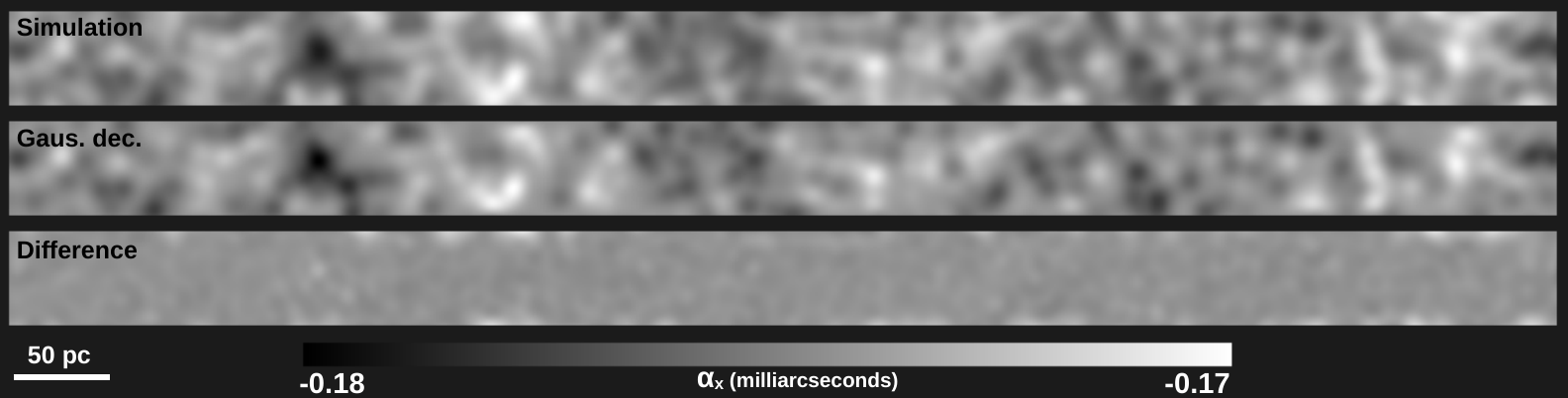}  
      \caption{Deflection field ($x$ component) from a simulation of $\psi$DM with $\lambda=10$\,pc. The top panel shows the original simulated deflection field. The middle panel illustrates the Gaussian decomposition of this field at 50 times the resolution and obtained with 500 Gaussian functions. The bottom panel represents the difference, which is at the percent level. We use the Gaussian decomposition to interpolate the deflection field to much higher resolutions. The Gaussian decomposition for the $y$ component of the deflection field looks similar.}
         \label{Fig_GaussianDecomposition}
\end{figure}
%%%%%%%%%%%%%%%%%%%%%%%%%%%%%%%%%%%%%%%%%%%%%%%%%%%%%%%%%%%

%The full simulation of $\psi$DM follows \cite{Amruth2023} {\bf Amruth, can you add a few lines here?}. We consider a model of $\psi$DM with $\lambda_{\rm dB}=10$\,pc corresponding to a massive cluster-scale lenses. In Figure~\ref{Fig_wDM} we show the effect of $\psi$DM over a small region in the observer plane. For this particular case the source is at $z=1$. We simulate regions at different distances to the CC by modifying macromodel magnifications, with values increasing toward the cluster CC. 

%%%%%%%%%%%%%%%%%%%%%%%%%%%%%%%%%%%%%%%%%%%%%%%%%%%%%%%%%%%
%Figure made Dell /home/jdiego/Lesnsing/Microlensing/WaveDM_Amruth/Magnification_ImagePlane_WaveDM_DifferentMu_POS.odp
%PNG figure in PDF made by /home/jdiego/Lesnsing/Microlensing/WaveDM_Amruth/Plot_Magnification_DifferentMu.pro
\begin{figure*} 
   \includegraphics[width=18.0cm]{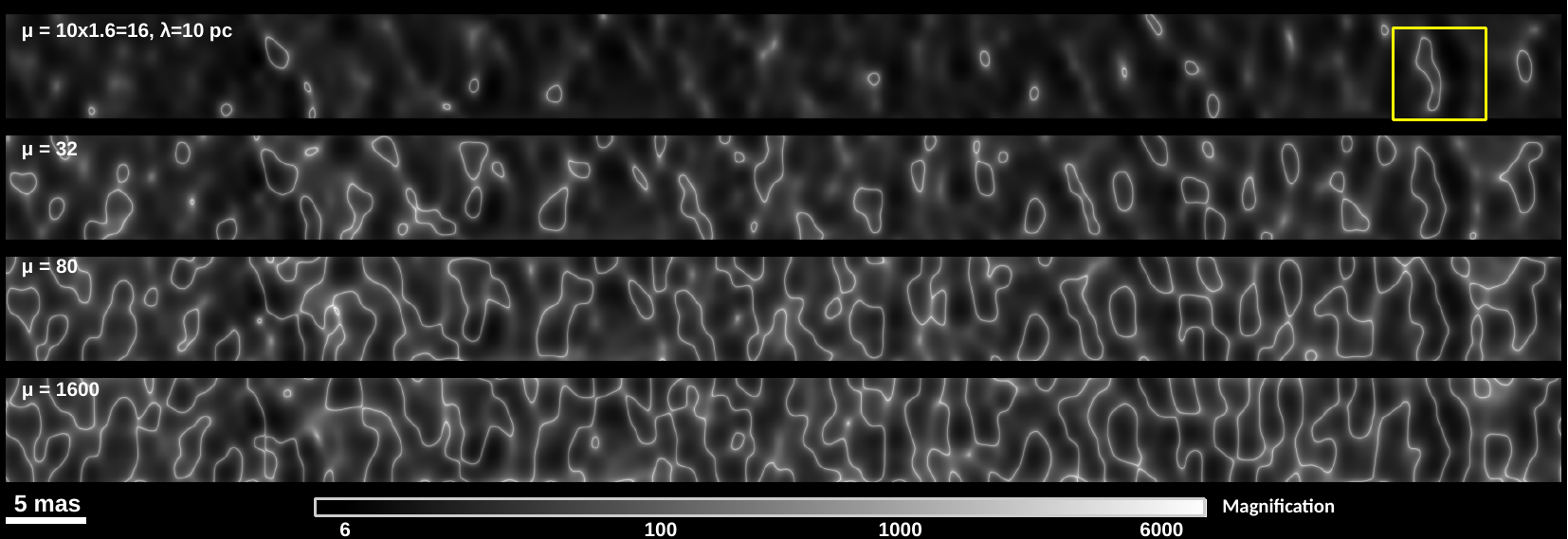} 
      \caption{Magnification in the image plane for a $\psi$DM model with a de Broglie wavelength of 10\,pc and in regions with positive parity. The top panel shows the $\psi$-CC (magnification in gray scale) formed by $\psi$DM with $\lambda=10$\,pc in a region with macromodel magnification $\mu=10 \times 1.6 = 16$ and positive parity. The bottom panels show the same realization of $\psi$DM but for increasing values of the tangential  macromodel magnification (in the horizontal direction) while the radial component (vertical direction) is maintained at $\mu_r=1.6$. In this configuration, and in the absence of $\psi$DM fluctuations  or small-scale perturbations in the deflection field, the cluster CC would appear as a single vertical line when $d=0$ or $\mu_t=\infty$. The scale is the same in all plots.  The corresponding distribution of magnification values is shown in Figure~\ref{Fig_Magnification_wDM_PDF}. The yellow square in the top panel indicates the region shown in Figure~\ref{Fig_Magnification_POS_NEG}.
         }
         \label{Fig_Magnification_wDM}
\end{figure*}
%%%%%%%%%%%%%%%%%%%%%%%%%%%%%%%%%%%%%%%%%%%%%%%%%%%%%%%%%%%

\section{Resolving caustics in $\psi$DM}\label{Sec_wDM_Sims}
%%%%%%%%%%%%%%%%%%%%%%%%%%%%%%%%%%%%%%%%%%%%%%%%%%%%
Simulations of the CCs and caustics produced by $\psi$DM have already been presented by \cite{Chan2020}, \cite{Amruth2023}, and \cite{Powell2023}. This pioneering work has shown how $\psi$DM can create an intricate corrugated network of CCs and caustics around the position of the lens CC or caustic, with small $\psi$-CC and $\psi$-caustics on scales comparable to the de Broglie wavelength. 
However, this earlier work lacks the resolution to resolve individual $\psi$-caustics with sufficient detail. Up to now, the best view of the caustic network produced by $\psi$DM was given by  \cite{Amruth2023}, who reach a resolution of 0.151 milliarcseconds per pixel in the image plane. 

In order to properly resolve the caustics and derive reliable probability distributions for the magnification in the source plane, the resolution needs to be increased significantly. At the same time, since we want to explore regions that are sufficiently close to the lens CC, where the macromodel magnification can easily surpass $\mu=1000$, a $\psi$DM simulation needs to cover an area sufficiently large in the image plane to be able to contain multiple $\psi$DM fluctuations. Such a simulation (large area at very high resolution) is computationally very demanding. Instead, we produce a simulation at lower resolution and rely on a model fit to the simulation to interpolate the deflection field to much higher resolution. We take advantage of the fact that the fluctuations in the deflection field of $\psi$DM are also very well described by Gaussian functions and construct a 2D model for the deflection field as a superposition of $N_G$ Gaussian functions, where we fix the scale of the Gaussian functions (to a size smaller than $\lambda_{\rm dB}$) and the only free parameters are their position and amplitude. We repeat this process for both deflection fields, $\alpha_x$ and $\alpha_y$.

We find that for our region of interest ($\sim 800 \times 50\, {\rm pc}^2$ in the image plane) with $N_G=500$ and $\sigma=0.6\lambda_{\rm dB}$, we can reach percent-level precision in the description of the simulated deflection field. Using the derived set of 500 Gaussian functions, we then recompute the deflection fields, $\alpha_x$ and $\alpha_y$, in the same region but with a resolution 50 times higher, that is with a pixel scale of 2 $\mu$arcsec per pixel. This approach is preferred over classic linear, cubic, or spline interpolation of the deflection field that can produce satisfactory results only for moderate increases in the resolution of a factor $<10$. An example of the performance of the Gaussian decomposition is shown in Figure~\ref{Fig_GaussianDecomposition}. Artifacts can be found in the edges of the region, but these are small and have no significant impact on our results. 

%%%%%%%%%%%%%%%%%%%%%%%%%%%%%%%%%%%%%%%%%%%%%%%%%%%%%%%%%%%
% Figure made by Dell Screenshot 
% Read files in /home/jdiego/Lesnsing/Microlensing/WaveDM_Amruth
% In IDL read the two files
%Magnification_All_50Kx2KPixel2muas_Mu10x1p6_WaveDM10pc_NEG.sav
%Magnification_All_50Kx2KPixel2muas_Mu10x1p6_WaveDM10pc_POS.sav
% Take the log10, and make the difference of CC maps. 
% Crop around pixels X=41525, Y=1326 with an area of 2600x260 pixels and TVSCL
\begin{figure} 
   \includegraphics[width=9.0cm]{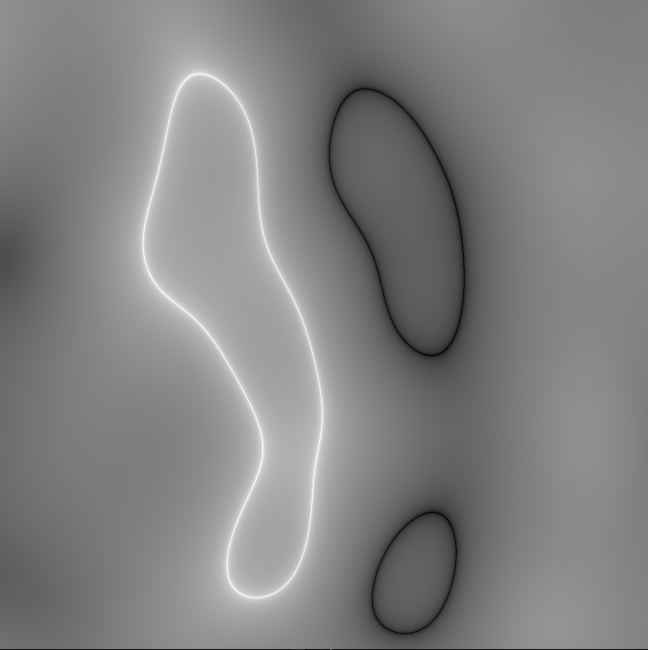}  
      \caption{Dependency of CCs with parity. The image shows the sum of the logarithm of magnifications (gray scale), ${\rm log}(\mu^{+})-{\rm log}(\mu^{-})$, from the same fluctuation but for the two parities, $\mu^{+}$ and $\mu^{-}$. The white line is the $\psi$DM-CC when the $\psi$DM fluctuation is on the side with positive parity. 
      The black line is the $\psi$DM-CC when the same  $\psi$DM fluctuation is on the side with negative parity. The $\psi$DM fluctuation for the case with positive parity is also highlighted with a yellow rectangle in Figure~\ref{Fig_Magnification_wDM}, where it can be seen how when the same fluctuation is on the side with negative parity, the $\psi$DM-CC forms around the low-magnification region of the positive parity magnification. 
      }
         \label{Fig_Magnification_POS_NEG}
\end{figure}
%%%%%%%%%%%%%%%%%%%%%%%%%%%%%%%%%%%%%%%%%%%%%%%%%%%%%%%%%%%

\section{Magnification of $\psi$DM in the image plane}\label{Sec_CC}
%%%%%%%%%%%%%%%%%%%%%%%%%%%%%%%%%%%%%%%%%%%%%%%%%%%%%%%%%
The magnification in the image plane is not always representative of the observed magnification, especially for $\psi$DM models, since in this case a source typically forms multiple images (usually separated by distances comparable to $\lambda_{\rm dB}$), all of them very close to each other and appearing in astronomical images as a single unresolved source. The observed flux is then the sum of all the fluxes from the multiple counterimages. Similarly, the observed magnification corresponds to the sum of all magnifications from the multiple images which is better estimated in the source plane, not in the image plane, since the source plane gives the total magnification of all images produced in the image plane. However, it is still instructive to study the properties of the magnification in the image plane, since some properties of the lensed sources can only be measured in the image plane and depend on the DM model. One such property is the clustering of the observed images, since these can form around the projected $\psi$DM constructive (or destructive) interference, hence revealing the scale and mass of the ALP.

In Figure~\ref{Fig_Magnification_wDM} we show the magnification (on a log scale) for a $\psi$DM model with $\lambda = 10$\,pc and in different portions of the image plane at different distances $d$ from the CC, where the macromodel magnification takes different values ($\mu=\mu_t\mu_r\propto d^{-1}$ and indicated in the top left of each panel). For simplicity we assume the radial component of the magnification is the same in all cases ($\mu_r=1.6$). This is well motivated since near the CC only the tangential component of the macromodel magnification changes rapidly. All regions shown in the figure are in a portion of the lens plane where images have positive parity. The CCs look very similar for the region on the opposite side of the CC and with negative parity. 

From top to bottom we see how small $\psi$-CCs form near the peaks of the $\psi$DM projected field. At relatively small macromodel magnification factors the CCs are not connected, but as we increase the macromodel magnification the $\psi$-CCs grow in size and start merging with neighboring $\psi$-CCs. At macromodel magnification factors $\mu=80$ the $\psi$-CCs almost fill the available space. At the highest macromodel magnification we consider, $\mu=1600$, the $\psi$-CCs have already formed a corrugated network of $\psi$-CCs. 

The appearance of the $\psi$-CCs and the corrugated network is very similar when computed assuming the macromodel magnification is negative (or on the side where macroimages have negative parity). Only in this case, $\psi$-CCs tend to form on the complementary space where $\psi$DM forms pockets of small magnification in Figure~\ref{Fig_Magnification_wDM}. An example is illustrated in Figure~\ref{Fig_Magnification_POS_NEG}, where we show the magnification in both scenarios of positive and negative parities, and for the same $\psi$DM fluctuation. The white lines correspond to the same $\psi$-CCs in the yellow square in Figure~\ref{Fig_Magnification_wDM} (positive parity), while the black curve is the $\psi$-CCs that form around the same region but when we change the macromodel magnification from $\mu=16$ to $\mu=-16$. This effect was already noted by \cite{Amruth2023}, who describe how when the parity is positive, $\psi$-CCs form around the positive fluctuations, while when the parity is negative, $\psi$-CCs form around the negative fluctuations. 
In this sense, the morphology of $\psi$-CC is different from those predicted by microlenses or millilenses that form a well-defined hourglass shape but which is oriented differently depending on whether the parity of the macromodel is positive or negative \citep{Chang1979}.

%%%%%%%%%%%%%%%%%%%%%%%%%%%%%%%%%%%%%%%%%%%%%%%%%%%%%%%%%%%
%Figure made by Dell /home/jdiego/Lensing/Microlensing/WaveDM_Amruth/Plot_Magnification_DifferentMu.pro 
\begin{figure} 
   \includegraphics[width=9.0cm]{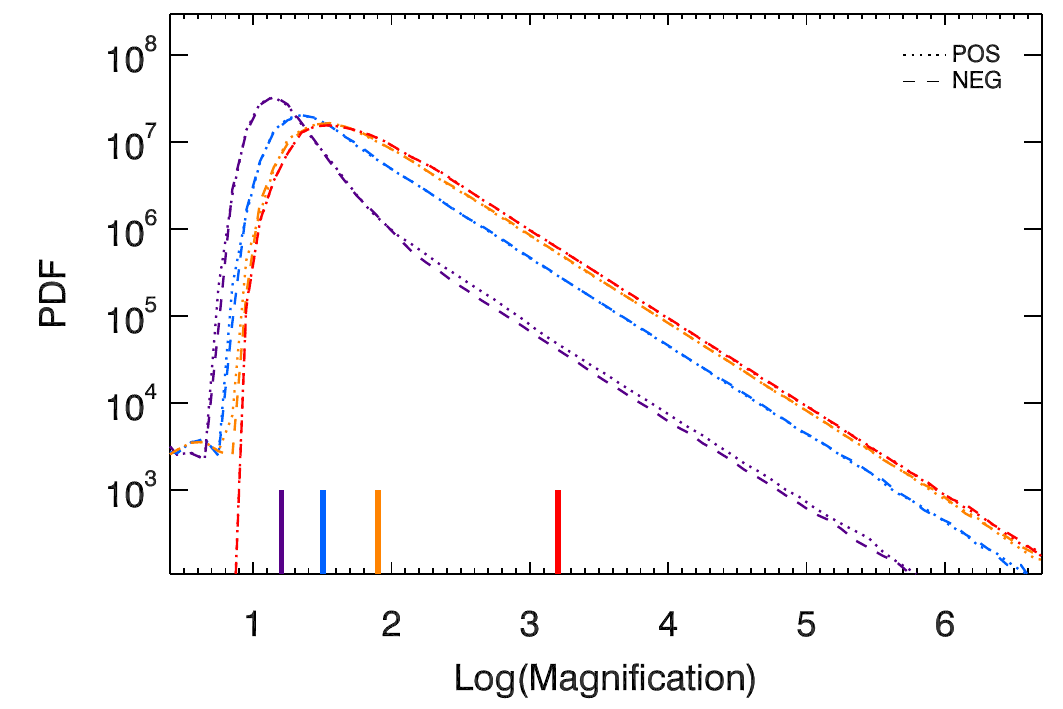}  
      \caption{Probability of magnification for $\psi$DM in the image plane. The probability is derived from the regions shown in Figure~\ref{Fig_Magnification_wDM}. Each color corresponds to a different macromodel magnification, indicated by the thick vertical solid lines. The dotted and dashed lines corresponds to the positive (POS) and negative (NEG) macromodel magnifications, respectively. The probability of magnification in the image plane is not representative of the observed magnification since it does not account for the multiplicty of images.}
         \label{Fig_Magnification_wDM_PDF}
\end{figure}
%%%%%%%%%%%%%%%%%%%%%%%%%%%%%%%%%%%%%%%%%%%%%%%%%%%%%%%%%%%

The probability of magnification computed in the image plane and for both parities is shown in Figure~\ref{Fig_Magnification_wDM_PDF}. Dotted lines indicate the relative probability of magnification (computed in logarithmic bins) for regions with varying macromodel magnification on the side with positive parity. Dashed lines show the case for the same regions but negative parity. Different colors are used to distinguish different macromodel values. These colors, together with the corresponding macromodel value, are given in the bottom of the figure as vertical solid line markers. The positions of these markers indicate the macromodel magnification for each particular region. For $\psi$DM on cluster lenses there seems to be no distinction between the sides with positive or negative parity \citep[but see][for galaxy-scale lenses where the situation is different]{Amruth2023}, making it a distinctive feature of this model when compared to CDM+substructure predictions \citep{Diego2018,Williams2024}. The probability of magnification in the image plane is very similar for macromodel magnification factors $\mu \gtrsim 100$. This would correspond to the saturation regime described by \cite{Diego2018} for the case of microlensing near CCs, but in this case applied to $\psi$DM, where from Figure~\ref{Fig_Magnification_wDM} we appreciate how the corrugated network has already formed for $\mu\approx100$. 

In all cases, the tail of the distribution follows the canonical $dN/d\mu \propto \mu^{-2}$, or $dN/dlog(\mu) \propto \mu^{-1}$, as in CDM. The probability of magnification shown in Figure~\ref{Fig_Magnification_wDM_PDF} does not include the effect of microlenses nor millilenses, but these are known to modify only slightly the probability of magnification when considering large areas ($A>> \lambda_{dB}^2$) in the image plane \citep{Diego2018,Williams2024}. On the other hand, on smaller scales, comparable to $\lambda_{dB}$, microlenses or millilenses are expected to modify the probability of magnification significantly, disrupting the $\psi$DM-CCs and redistributing the magnification lowering the magnification at the $\psi$DM-CC positions, and increassing it around microlenses and millilenses. A detailed study of this effect on very small scales is beyond the scope of this paper and will be treated in a future work.

%%%%%%%%%%%%%%%%%%%%%%%%%%%%%%%%%%%%%%%%%%%%%%%%%%%%%%%%%%%
%Figure made Dell /home/jdiego/Lesnsing/Microlensing/WaveDM_Amruth/Caustics_WaveDM_DifferentMu.odp
%PNG figure in PDF made by /home/jdiego/Lesnsing/Microlensing/WaveDM_Amruth/Plot_Caustics_MakePDFMu_DifferentMu.pro 
\begin{figure*} 
   \includegraphics[width=18.0cm]{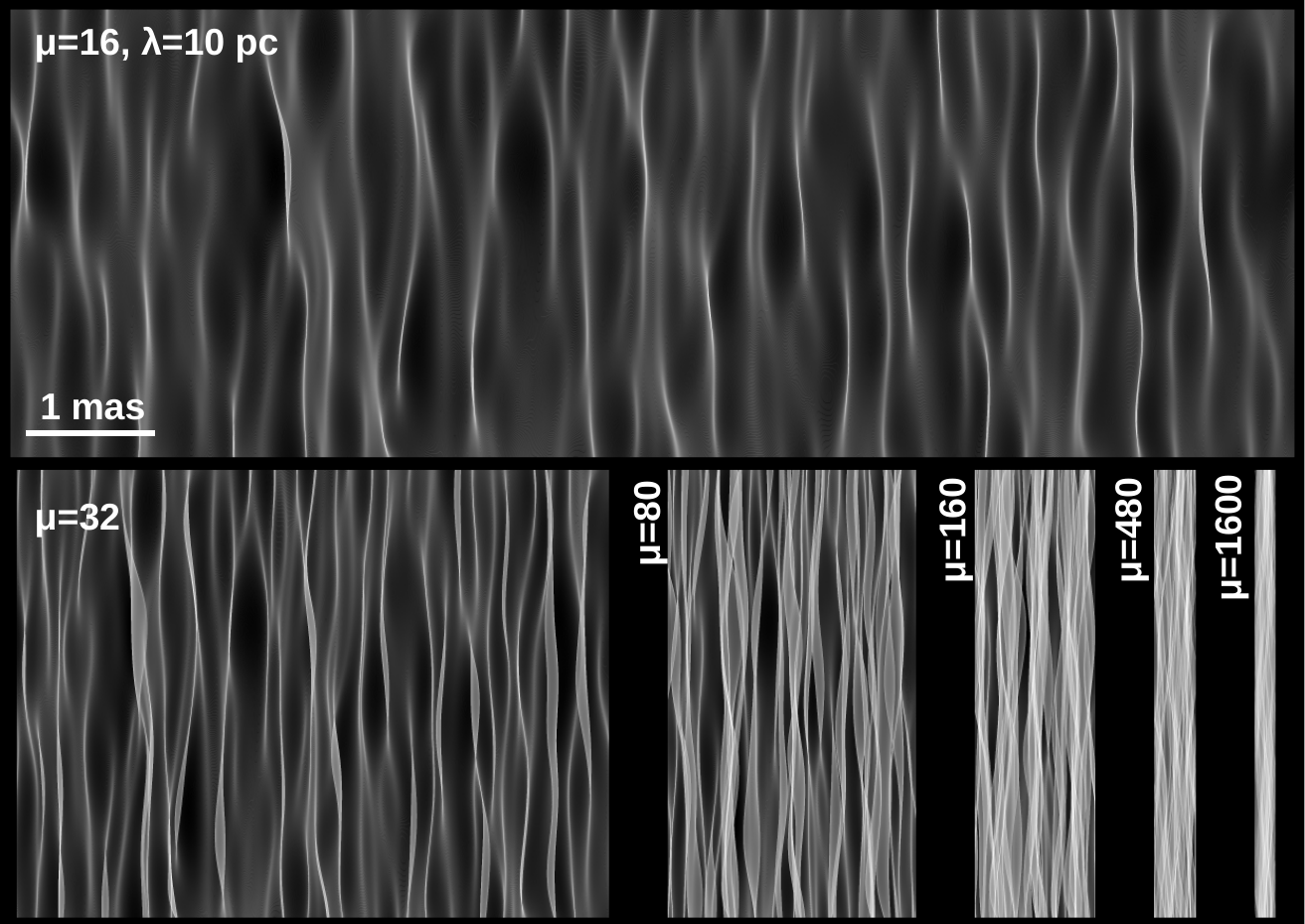} 
      \caption{Caustics for $\psi$DM. The top panel shows the caustics (magnification in gray scale) formed by $\psi$DM with $\lambda=10$\,pc in a region with macromodel magnification $\mu=10\times1.6=16$ and positive parity. The bottom panels display the same realization of $\psi$DM but with increasing values of the tangential macromodel magnification while the radial component is maintained at $\mu_r=1.6$. The cluster caustic is not shown would be a thin vertical line to the right of the bottom-right panel. The scale is the same in all plots. The caustics with the same macromodel magnification but negative parity look almost indistinguishable. The corresponding distributions of magnification values are shown in Figure~\ref{Fig_Caustics_wDM_PDF}
         }
         \label{Fig_Caustics_wDM}
\end{figure*}
%%%%%%%%%%%%%%%%%%%%%%%%%%%%%%%%%%%%%%%%%%%%%%%%%%%%%%%%%%%

\section{Magnification of $\psi$DM in the source plane}\label{Sec_Caustics}
%%%%%%%%%%%%%%%%%%%%%%%%%%%%%%%%%%%%%%%%%%%%%%%%%%%%%%%%%
The true magnification experienced by a small source moving across the web of $\psi$-caustics is the one computed in the source plane. We compute the magnification in the source plane by standard ray tracing and for the same regions with varying macromodel magnification discussed in the previous section.

The resulting magnification pattern in the source plane is displayed in Figure~\ref{Fig_Caustics_wDM}. This image represents the best view of the $\psi$DM-caustics to date. The top panel shows the region with the smallest macromodel magnification ($\mu=16$). The bottom row illustrates the remaining regions with increasing macromodel magnification values (shown in, or near, the top-left corner of each panel). The cluster CC is not shown but in our configuration it would appear as a vertical white line at the right edge of the bottom panel. This figure shows only the case of the positive parity; the negative parity looks very similar and hence is not displayed here.

\begin{figure} 
   \includegraphics[width=9.0cm]{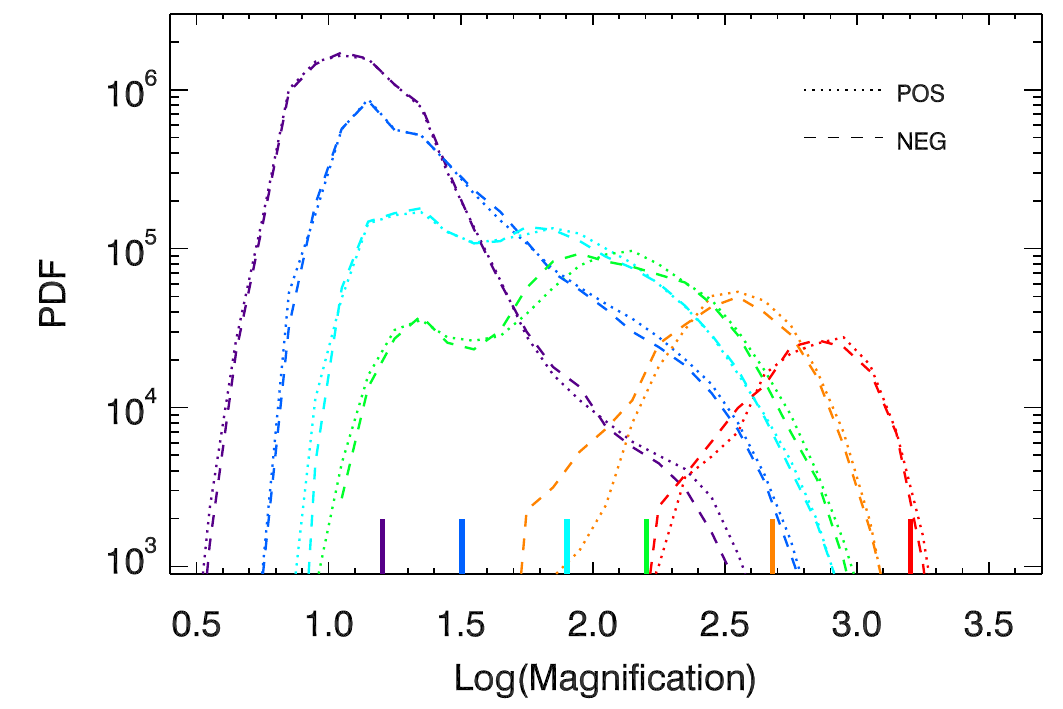}  
      \caption{Probability of magnification for $\psi$DM in the source plane. The probability is derived from the regions shown in Figure~\ref{Fig_Caustics_wDM}. Each color corresponds to a different macromodel magnification, indicated by the thick vertical solid lines. The dotted and dashed lines corresponds to the positive and negative macromodel magnifications respectively. }
         \label{Fig_Caustics_wDM_PDF}
\end{figure}

One striking difference between the $\psi$DM and CDM substructures (microlenses and millilenses) is that in the case of  $\psi$DM the magnification pattern is nearly identical, independent of the parity. This was already demonstrated by the toy model discussed in Section~\ref{Sec_wDM}. 
This similarity is better demonstrated in Figure~\ref{Fig_Caustics_wDM_PDF}, where we show the distribution of magnification in the source plane for regions with different parity and different macromodel magnification values. The line and color scheme is similar to that in Figure~\ref{Fig_Magnification_wDM_PDF}. 

This is not the case  in classic CDM simulations where substructure is composed of either point sources (microlenses) or cuspy halos (millilenses). These halos correspond to dwarf galaxies, globular clusters, ultracompact dwarf galaxies, or DM halos in general. One common characteristic of the microlenses and millilenses studied in the context of CDM is that when they are found in regions of negative parity they can significantly demagnify sources below the macromodel value, while on the side with positive parity only very modest demagnification can take place. 
In CDM, significant demagnification on the side with positive parity would be possible only with objects having negative mass. These hypothetical objects behave as similar objects but with positive mass on the side with negative parity. In $\psi$DM, negative-mass objects are equivalent to the negative $\delta\rho/\rho<<1$ fluctuations which can naturally demagnify distant objects aligned behind them.  

In the $\psi$DM scenario, demagnification can take place for both parities, and with similar probability independent of the parity. On the other hand, similar to the case of CDM plus substructure, the maximum variance in the magnification takes place in situations where the saturation regime is reached, which in our case is for macromodel magnification $\mu\approx 100$.

For the cases with macromodel magnification $\mu<100$, the tail of the distribution in Figure~\ref{Fig_Caustics_wDM_PDF} falls faster than the expected $dN/dlog(\mu)\propto \mu^{-2}$, but this is partially an artifact due to the limited size of the pixel in the simulation. We do expect the $dN/dlog(\mu)\propto \mu^{-2}$ tail to extend to higher magnifications, but not to infinity because at some point microlenses near the $\psi$-caustics (and not included in our simulation) will transform the $\mu^{-2}$ power law into log-normal as described by \cite{Diego2018,Diego2020} and \cite{Palencia2023}. For large macromodel values $\mu>>100$, we observe the log-normal distribution predicted also for the case of microlenses near CCs. In addition, a finite-pixel effect is expected here for the largest magnification factors, but combined with the faster decline from the log-normal  behavior due to the combined effect of  overlapping $\psi$-caustics. Adding microlenses to the $\psi$DM model in the regime where the macromodel magnification is already $\mu>>100$ should have little impact on the log-normal distribution of magnifications, since the log-normal shape (and position of the central peak) would be maintained, but the addition of microlenses should broaden it by a small amount. A similar argument can be made for the addition of millilenses, except in this case we expect a modification of the probability of magnification near the millilens that can imprint a distortion in the deflection field on a scale similar to the de Broglie wavelength. We leave the detailed study of $\psi$DM plus millilenses and microlenses for  future work.  The case of microlenses plus microlenses was studied (in the context of CDM) by \cite{Diego2024b}.

If we look at the maxima of all the probabilites in Figure~\ref{Fig_Magnification_wDM_PDF}, we observe how these maxima scale as $1/\mu$, departing from the expected $1/\mu^2$ of the macromodel (when taking logarithmic bins in the magnification, as done for this figure). This is because the area in the image plane is the same for all curves, but in a real-world scenario, the curves with lower macromodel magnification would originate from regions with a correspondingly larger area, so this envelope should be corrected by an additional factor $\mu^{-1}$, and hence recovering the expected $\mu^{-2}$ scaling of the macromodel. In other words, as in the case of substructure in CDM, $\psi$DM does not modify the overall probability of magnification on scales larger than the de Broglie scale. Instead, the magnification (or flux) is simply redistributed, with $\psi$DM fluctuations far away from the cluster CC borrowing high magnification values from the CC.  This mechanism is similar in CDM+substructure. 

As in CDM, we do expect the maximum magnification of a small object (for instance a star) in the corrugated network to be significantly less than in the case of smooth CDM. This reduction in maximum magnification is due to the small-scale perturbations in the deflection field, which increase the curvature of the deflection field on small scales. Since the maximum magnification when crossing a caustic is inversely proportional to this curvature \citep{Schneider1992}, $\psi$DM fluctuations near the cluster CC also decrease the maximum magnification of a star when crossing a $\psi$DM-caustic. The amount of this reduction is expected to scale inversely proportional to $\lambda_{\rm dB}$, since smaller $\lambda_{\rm dB}$ results in higher local curvature. The exact scaling of the maximum magnification with $\lambda_{\rm dB}$ is also beyond the scope of this paper and will be treated in  future work. 

%%%%%%%%%%%%%%%%%%%%%%%%%%%%%%%%%%%%%%%%%%%%%%%%%%%%%%%%%%%
%Figure made by Dell /home/jdiego/Lesnsing/Microlensing/WaveDM_Amruth/Plot_Caustics_MakePDFMu_DifferentMu.pro 
%%%%%%%%%%%%%%%%%%%%%%%%%%%%%%%%%%%%%%%%%%%%%%%%%%%%%%%%%%%

\section{Discussion}\label{sect_Disc}
\label{Sec_Discussion}
%%%%%%%%%%%%%%%%%%%%%%%%%%%%%%%%%%%%%%%%%

Perhaps the most interesting feature of the curves shown in Figure~\ref{Fig_Caustics_wDM_PDF} is the fact that $\psi$DM perturbations on the side with positive parity have the capacity to demagnify significantly more than substructures in CDM (such as millilenses or microlenses). To better show this difference, we can compare the expected probability of magnification in CDM plus substructure with the $\psi$DM case. We do this by adding microlenses to a smooth CDM model, since microlenses, similarly to $\psi$DM fluctuations, are ubiquitous in the image plane. Millilenses, on the other hand, although producing deflection fields with similar amplitude as $\psi$DM, are much more rare in the  image plane, so the probability of magnification depends on whether it is computed near a millilens or far away from it. 

\begin{figure} 
   \includegraphics[width=9.0cm]{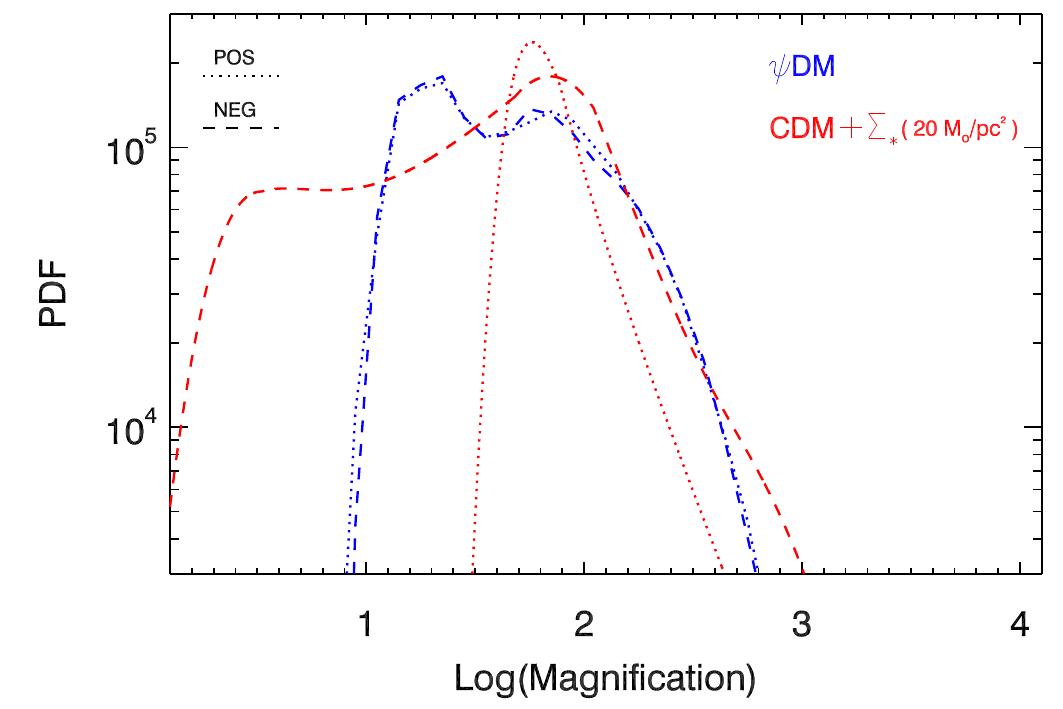}  
      \caption{Comparison of magnification with $\psi$DM and CDM+microlenses. Dotted lines represent regions having positive parity while dashed lines are for negative parity. In all curves, the macromodel magnification is set to 80. In the case of CDM (red curves), the substructure is created by microlenses with surface mass density 20\,M$_{\odot}\,{\rm pc}^{-2}$. The addition of microlenses to $\psi$DM would broaden the dashed line curves a bit but maintaining the demagnification peak on the side with positive parity.}
         \label{Fig_wDM_vs_cDM}
\end{figure}

We consider the particular case of macromodel magnification $\mu_{macro}=80$ (light-blue curves in Figure~\ref{Fig_Caustics_wDM_PDF}) for which a prominent demagnification peak forms around $\mu=20$. This value of $\mu_{\rm macro}$ corresponds to the threshold where the corrugated network of $\psi$-caustics starts to form and magnification distortions are greatest. 
For the microlenses we consider a case with surface mass density of microlenses $\Sigma_{*} \approx 20\, {\rm M}_{\odot}\, {\rm pc}^{-2}$. This is a value typically found near CCs, where stellar microlenses from the intracluster medium contribute $\sim 1\%$ to the total projected mass, and this is typically $\Sigma_{\rm Tot} =\kappa \times\Sigma_{\rm crit} \approx 2000\, {\rm M}_{\odot}\, {\rm pc}^{-2}$ (for convergence values $\kappa\approx 0.6$, as usually found near cluster CCs). We compute the probability of magnification in the case of CDM+microlenses with the microlensing code from \cite{Palencia2023} adopting the parameters described above. The resulting probability of magnification in both cases, $\psi$DM and CDM+microlenses, is shown in Figure~\ref{Fig_wDM_vs_cDM}. 
The probability of magnification for $\psi$DM is indicated as dark blue curves (the same curves are  light blue in Figure~\ref{Fig_Caustics_wDM_PDF}). The curves for CDM+microlenses are shown as red curves. When comparing the two curves with positive parity, at low magnification values $\mu\approx 20$, $\psi$DM (blue dotted) has orders of magnitude more probability of demagnifying than the  CDM+microlenses case (red dotted). Increasing the amount of microlenses would result in a slightly wider curve, but not sufficient to significantly increase the probability of relative demagnification.

Adding microlenses to $\psi$DM does not change this conclusion. Microlenses would widen the dark- blue dotted line as if convolved by the red dotted line, still leaving a prominent peak at $\mu\approx 20$, which cannot be reproduced by CDM plus substructure (even millilenses). Demagnification of images by more than a factor of $\sim 2$ is only possible on the side with positive parity for $\psi$DM. This demagnification is not only possible in $\psi$DM, but very likely at the boundaries of the corrugated network. 
On the side with negative parity, similar high magnification factors and probabilities are expected in $\psi$DM and CDM+substructure. In contrast, demagnification is less pronounced in the case of $\psi$DM than in the case of CDM+substructure, so demagnification factors greater than one order of magnitude are much less likely in $\psi$DM, while they are still quite likely in CDM+substructure. 

When confronting these predictions with observations, a good example to consider is the Ly-C knot in the Sunburst arc \citep{Rivera-Thorsen2019}, where up to 12 highly magnified images of the same compact structure are seen multiply lensed. This Ly-C knot is sufficiently large to be insensitive to microlensing, but is small enough that it can be affected by $\psi$DM perturbations. Other small structures in the arc are also observed multiple times, with some of them (such as the alleged lensed star Godzilla) presenting highly anomalous magnifications \citep{DiegoGodzilla}. 
%Detailed lens modelling of this arc can reveal the need for small-scale substructure not consistent with CDM. In some cases, this substructure may be well explained in the context of the standard CDM, as for Godzilla where \cite{DiegoGodzilla} claimed that an undetected dwarf-galaxy millilens was responsible for the anomalous magnification of Godzilla. The alleged millilens was later discovered in deeper JWST data by \citep{Choe2024}. In other cases, 
These authors suggested that Godzilla can be explained in the context of the standard CDM, and claimed that an undetected dwarf-galaxy millilens was responsible for its anomalous magnification. The alleged millilens was later discovered in deeper JWST data by \citep{Choe2024}. While CDM may be the right explanation at this point, if more Godzilla-like objects are detected, CDM may become disfavored because the amount of substructure needed (or its ubiquity) may be inconsistent with the CDM scenario and be better explained by $\psi$DM, where lensing perturbations are expected to be affecting all images in the Sunburst arc.
The very clear and fundamental differences in how images with different parities can be demagnified can be used to discriminate between CDM and $\psi$DM using current observations and detailed lens models of arcs such as the Sunburst arc. 

Another interesting object is the Dragon arc in Abell 370, where \cite{Fudamoto2024} reports over 40 alleged microlensing events. Most of these events are found on the side of the CC with negative parity as shown by \cite{Broadhurst2024}. This is consistent with the findings in this paper where before adding microlenses, the probability of magnification is similar on both sides of the CC, but when microlenses are added images with negative parity are more likely to be detected since they carry larger magnifications than their counterimages with negative parity, as a consequence of the demagnification produced by microlenses that needs to be compensated with higher magnification factors. We leave the detailed interpretation of these events in the context of $\psi$DM for a future paper, after the macromodel around the Dragon arc has been updated with the new constraints from the recent JWST observations. 

%%%%%%%%%%%%%%%%%%%%%%%%%%%%%%%%%%%%%%%%%%%%%%%%%%%%%%%%%%%
%Figure made by Dell /home/jdiego/Lesnsing/Microlensing/WaveDM_Amruth/Plot_Caustics_MakePDFMu_DifferentMu.pro 
%%%%%%%%%%%%%%%%%%%%%%%%%%%%%%%%%%%%%%%%%%%%%%%%%%%%%%%%%%%

An additional way of distinguishing between CDM and $\psi$DM is based on the scale of the background source. As we have seen, both $\psi$DM and microlenses produce ubiquitous fluctuations in the deflection field, though they result in different probabilities of demagnification. Very small but luminous objects, such as supergiant stars, can be very effectively magnified and demagnified by microlenses. The size of microcaustics from stars in the intracluster medium is typically $10^{-3}$\,pc, so objects 1\,pc in size become immune to microlensing distortions and can be magnified or demagnified only by more massive structures in CDM, such as millilenses. At 1\,pc, $\psi$DM is still very efficient at magnifying and demagnifying, and at $\sim 0.5$\,pc the efficiency is maximum for de Broglie wavelengths $\lambda \approx 10$\,pc (see Figure~\ref{Fig_Caustics_wDM_PDF}). A high occurrence of demagnification of parsec-scale structures would be inconsistent with CDM since the number density of millilenses in the image plane is not very high, while $\psi$DM fluctuations are always taking place in the image plane so the probability of demagnification is always $\sim 50\%$. 

Recent observations with JWST have revealed the existence of very small but bright star clusters at high redshift \citep{Vanzella2023,AngelaNature2024}, corroborating the highly compact nature of some of these luminous objects in local measurements \cite{Brown2021}. Compact stellar clusters make perfect candidates to map DM substructure in the lens plane on sub-parsec scales, and in particular to test $\psi$DM models. These objects are big enough that they should be almost insensitive to microlensing distortions, but sufficiently small to be smaller than the demagnification region in the source plane produced by $\psi$DM fluctuations. At the same time, these small clusters are  bright enough to still be observed with demagnification factors of 4--5 relative to the macromodel value, when $\mu_{\rm macro}\approx 100$, which is the range of macromodel magnifications where we expect $\psi$DM lensing distortions to be maximum. The spatial distribution of JWST microlenses in critically lensed galaxies recently uncovered by \cite{Fudamoto2024} and \cite{Yan2023} can help discriminate between these possibilities. Most of the new microlensing events are found in broad band around the CC, with a tendency toward negative parity. This is consistent with the $\psi$DM scenario once the extra magnification from microlenses, in addition to the boost in magnification provided by $\psi$DM fluctuations, is taken into account \citep[see][for details]{Broadhurst2024}.

So far the conclusions derived in this work are obtained for a $\psi$DM model with $\lambda_{\rm dB}=10$\,pc. The properties of magnification should depend on the de Broglie wavelength as indicated by earlier work \cite{Amruth2023}. A detailed study of the dependence of our conclusions on $\lambda_{\rm dB}$ is beyond the scope of this paper, but from earlier work we can conclude that the width of the corrugated network grows as $\lambda_{\rm dB}^{0.5}$, and in the limit of high boson mass (very small $\lambda_{\rm dB}$) one should recover the smooth-lens case with no corrugated network. It is expected that, as in the case of microlenses, the probability of magnification in $\psi$DM  can be entirely described by an effective surface mass density of $\psi$DM perturbations, $\Sigma_{\rm eff} = \Sigma_{\rm dB}(\lambda_{\rm dB})\times\mu_{\rm macro}$, where  $\Sigma_{\rm dB}(\lambda_{\rm dB})$ is some function of $\lambda_{\rm dB}$ of the form $\Sigma_{\rm dB}(\lambda_{\rm dB}) \propto \lambda_{\rm dB}^{\alpha}$ with $\alpha>0$ \cite[see][for further details]{Broadhurst2024}.

\section{Conclusions}
\label{Sec_Conclusions}
%%%%%%%%%%%%%%%%%%%%%%%%%%%%%%%%%%%%%%%%%
We have generated the first high-resolution images of the source plane in a model of $\psi$DM with $\lambda=10$\,pc and in regions very close to the CC of a galaxy cluster lens. The perturbations in $\psi$DM result in pockets of low magnification in the source plane with scales  of $\sim 0.5$\,pc. These low-magnification regions are typical in CDM scenarios plus substructure for images with negative parity, but we find that in the case of $\psi$DM they also appear on the side of the CC with positive parity. Demagnification of pc-scale objects in their positive-parity counterimages cannot be reproduced by substructure in CDM, but are likely in $\psi$DM making this a unique testable feature of $\psi$DM models. 

Images with positive and negative parity have very similar distributions of magnification. This differs distinctly from CDM predictions, where the positive- and negative-parity images are clearly distinguishable, especially for macromodel images with $\mu_{\rm macro}<100$. In particular, $\psi$DM alone cannot demagnify negative-parity images as much as CDM substructure. Adding microlenses to $\psi$DM increases the probability of demagnification for images with negative parity, bringing the results close to CDM predictions, but does not enhance significantly the demagnification of images with positive parity. This behavior is unique to $\psi$DM. Current and future observations of multiple images from parsec-scale star clusters near cluster CCs can be used to further test $\psi$DM models. 

\begin{acknowledgements}
%%%%%%%%%%%%%%%%%%%%%%%%%
 
 J.M.D. acknowledges support from project PID2022-138896NB-C51 (MCIU/AEI/MINECO/FEDER, UE) Ministerio de Ciencia, Investigaci\'on y Universidades.  
 J.L., S.K.L., and A.A. are funded by grants RGC/GRF 17312122 and RGC/CRF C6017-20G.
 R.A.W. acknowledges NASA JWST Interdisciplinary Scientist grants NAG5-12460, NNX14AN10G, and 80NSSC18K0200 from GSFC. 
 A.V.F. received financial assistance from the Christopher R. Redlich Fund and many individual donors. 
 L.L.R.W. acknowledges the support of HST SNAP program 17504.
 %P.L.K acknowledges NSF AAG 2308051.

\end{acknowledgements}

\bibliographystyle{aa} %use aa.bst
\bibliography{MyBiblio} % References in MyBiblio.bib run bibtex after latex/pdflatex 

%%%%%%%%%%%%%%%%%%%%%%%%%%%%%%%%%%%%%%%%%%%%%%%%%%%%%%%%%%%%%%%%%

\end{document}